\documentclass[reprint,amsmath,amssymb,aps,prl,nobibnotes,superscriptaddress]{revtex4-1}
\usepackage{graphicx}
\usepackage{dcolumn}
\usepackage{amsmath}
\usepackage{bm}
\usepackage[utf8]{inputenc}
\usepackage{mathtools}
\usepackage{pgffor}
\usepackage[normalem]{ulem}
\usepackage{xcolor}
\usepackage{BOONDOX-cal}
\usepackage{braket}
\usepackage{etaremune}
\usepackage{pdfpages}
\usepackage{BOONDOX-cal}

\makeatletter

\AtBeginDocument{\let\LS@rot\@undefined}
\makeatother

\newif\ifarXiv
\arXivtrue 

\begin{document}

\title{Ultraclean two-dimensional hole systems with mobilities exceeding 10$^7$ cm$^2$/Vs}

\author{Adbhut Gupta}
\email{Corresponding author: adbhutg@princeton.edu}
\affiliation{Department of Electrical and Computer Engineering, Princeton University, Princeton, NJ, USA}
\author{C. Wang}
\affiliation{Department of Electrical and Computer Engineering, Princeton University, Princeton, NJ, USA}
\author{S.K. Singh}
\affiliation{Department of Electrical and Computer Engineering, Princeton University, Princeton, NJ, USA}
\author{K.W. Baldwin}
\affiliation{Department of Electrical and Computer Engineering, Princeton University, Princeton, NJ, USA}
\author{R. Winkler}
\affiliation{Department of Physics, Northern Illinois University, DeKalb, Illinois, USA}
\author{M. Shayegan}
\affiliation{Department of Electrical and Computer Engineering, Princeton University, Princeton, NJ, USA}
\author{L.N. Pfeiffer}
 \affiliation{Department of Electrical and Computer Engineering, Princeton University, Princeton, NJ, USA} 
\date{\today}
\begin{abstract}

Owing to their large effective mass, strong and tunable spin-orbit coupling, and complex band-structure, two-dimensional hole systems (2DHSs) in GaAs quantum wells provide rich platforms to probe exotic many-body physics, while also offering potential applications in ballistic and spintronics devices, and fault-tolerant topological quantum computing. We present here a systematic study of molecular-beam-epitaxy grown, modulation-doped, GaAs (001) 2DHSs where we explore the limits of low-temperature 2DHS mobility by optimizing two parameters, the GaAs quantum well width and the alloy fraction ($x$) of the flanking Al$_x$Ga$_{1-x}$As barriers. We obtain a breakthrough in 2DHS mobility, with a peak value $\simeq 18 \times 10^6$ cm$^2$/Vs at a density of 3.8 $\times$ 10$^{10}$ /cm$^{2}$, implying a mean-free-path of $\simeq 57 \mu$m. Using transport calculations tailored to our structures, we analyze the operating scattering mechanisms to explain the non-monotonic evolution of mobility with density. We find it imperative to include the dependence of effective mass on 2DHS density, well width, and $x$. We observe concomitant improvement in quality as evinced by the appearance of delicate fractional quantum Hall states at very low density.
\end{abstract}
\maketitle

\section{Introduction}
The invention of the modulation-doping technique \cite{Dingle1978} in GaAs/AlGaAs heterostructures stands as a pivotal breakthrough in the material science and physics of two-dimensional (2D) carrier systems. Exponentially suppressing the inimical Coulomb scattering from intentional dopants, it opened exciting avenues for exploring physics in semiconductor systems with low disorder. Forming nearly perfect crystals when grown using molecular beam eptaxy (MBE), these heterostructures boast exceptional quality, as evidenced by record 2D mobility values \cite{Chung2021elec,Chung2022elec, Kulah2021, Manfra2014}. 

A key utilization of high-mobility GaAs 2D carrier systems is the investigation of exotic, many-body states arising from strong carrier-carrier interaction \cite{Shayegan2005, Halperin2020}. While 2D electron systems (2DESs) have long been at the forefront for exploration of interaction-driven phenomena, 2D \textit{hole} systems (2DHSs) offer an attractive alternative. At very low temperatures when the thermal energy is minimal, the strength of interaction is characterized by the relative strength of Coulomb energy ($E_C$) with respect to other energy scales such as Fermi ($E_F$) and cyclotron energies ($E_\textrm{cyc}$). At zero magnetic field ($B$), the relevant dimensionless parameter is $r_s = E_C/E_F$~$\propto$~$m^*/\sqrt{p}$  where $m^*$ is the effective mass and $p$ is the 2D density. At a given $p$, 2DHSs can have a much larger $m^*$ \cite{ Zhu2007, Watson2012} sometimes exceeding the free electron mass $m_e$, as compared to their electron counterparts ($m^*=0.067m_e$), enhancing the many-body effects. A notable example is the observation of a quantum Wigner crystal at zero $B$ in a dilute 2DHS \cite{Yoon1999}. At high $B$, the relevant interaction parameter is the Landau level (LL) mixing parameter, $\kappa=E_C/E_\textrm{cyc} \propto$~$m^*$ and leads to exotic phases such as Wigner crystal at relatively large LL filling factor $\nu$ \cite{Santos19921/3}, and even-denominator fractional quantum Hall states (FQHSs) \cite{Wang2022}. Very recently, numerous new even-denominator FQHSs were observed in high-mobility 2DHSs, for example at $\nu=$ 3/4, 3/8, 3/10 and 1/4 \cite{Wang2022, Wang2023A, Wang2023B}. Even-denominator FQHSs have garnered attention because they are expected to host quasiparticles obeying non-Abelian statistics \cite{Nayak2008}. This renders 2DHSs as possible contenders for fault-tolerant topological quantum computing. It is worth emphasizing that the above even-denominator FQHSs in the ultrahigh-quality 2DHSs are observed in the lowest ($N = 0$) LL ($\nu<1$), in contrast to the vast majority of even-denominator FQHSs in different materials which are reported in the excited ($N=1$) LL \cite{Nayak2008, Willett1987, Ki2014, Falson2015, Hossain2018, Shi2020, Dutta2022, KHuang2022}. Numerous other, strongly-correlated, many-body phases have also transpired in GaAs 2DHSs including bilayer FQHSs, bubble, and striped phases \cite{Manoharan1994signatures,Tutuc2004,Koduvayur2011,Kamburov2013,Liu2014,Liu2014even,Mueed2015,Liu2016,Jo2017,Ma2020,Ma2022,
Wang2023anisotropic}.

\begin{figure*}[!t]
\begin{center}
\includegraphics[width=1 \textwidth]{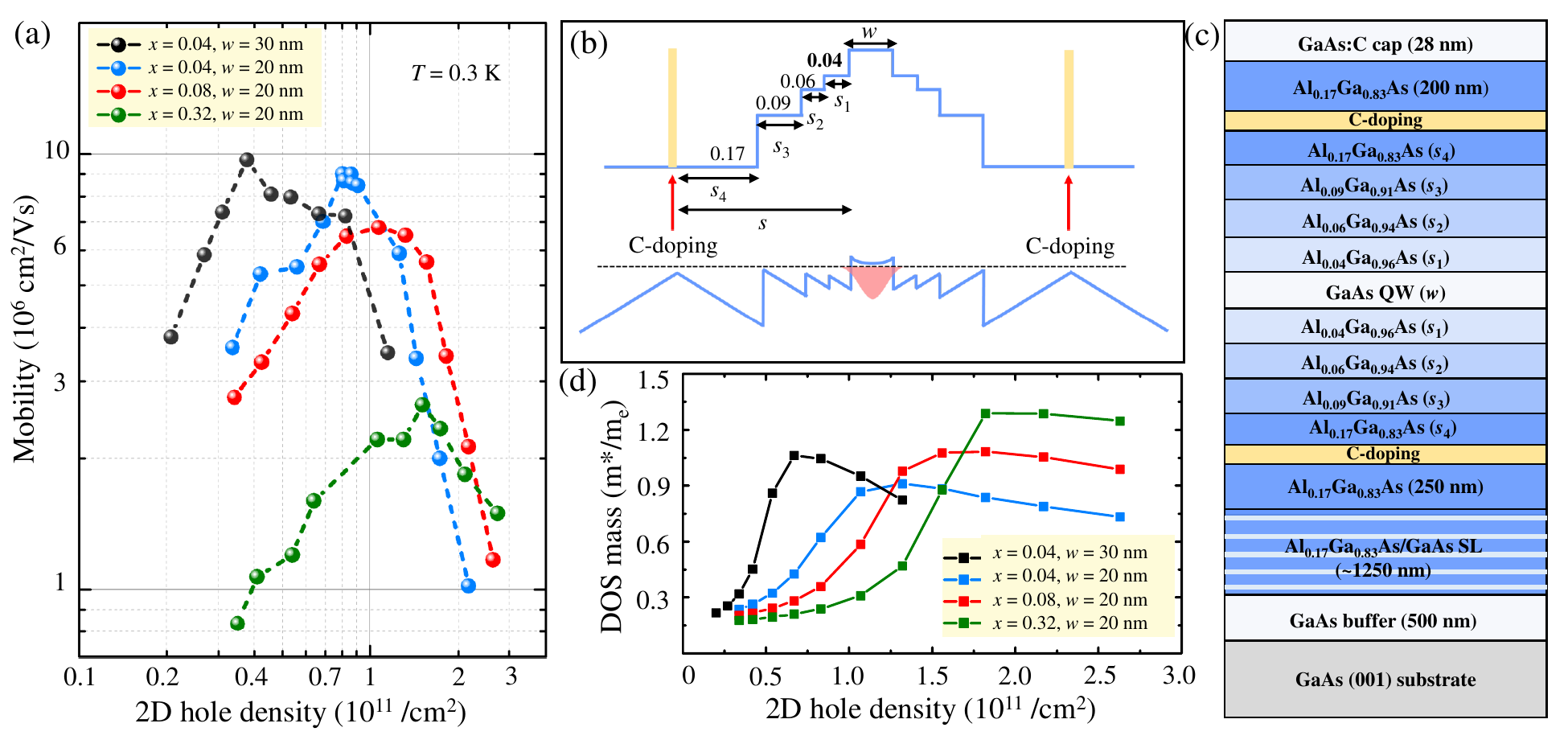}
\caption{\label{fig:fig1}(a) Hole mobility ($\mu$) plotted as a function of 2D density ($p$) for $w=$ 20 nm and $x=$ 0.32, 0.08, 0.04, and $w=$ 30 nm, $x=$ 0.04. (b) Schematic of the valence band for the $x=$ 0.04 design; top and bottom panels depict the valence band before and after hole charge transfer, respectively. Starting from $x=$ 0.17 near the doping, $x$ is lowered in steps to 0.09, 0.06 and 0.04. The step thicknesses ($s_1$, $s_2$, $s_3$, $s_4$) are carefully chosen to avoid parallel channels to form near the steps. Table S1 provides values of thicknesses for selected samples \cite{Supp}. The densities are tuned by varying the total spacer thickness ($s=s_1+s_2+s_3+s_4$). (c) Layer structure for the $x=$ 0.04 design. Layer structures for  $x=$ 0.08 and 0.32 designs are shown in Fig.~S1 \cite{Supp}. (d) Calculated density-of-states (DOS) effective masses $m^*$ in units of free electron mass $m_e$, vs $p$ for cases presented in (a).}
\end{center}
\end{figure*}

 In addition, strong spin-orbit (SO) coupling and heavy-hole light-hole mixing in the valence band enrich the physics of 2DHSs \cite{Winkler2003} as they cause non-linear LLs with several crossings \cite{Liu2014even, Liu2016, Ma2022}. These crossings can lead to interesting physics and can be tuned to create novel many-body ground states \cite{Wang2023anisotropic, Liu2014even, Ma2022}. Moreover, the valence band in GaAs consists of p-like atomic orbitals which reduces the overlap between the hole wavefunction and the nuclei, weakening the hyperfine interaction. This, along with the strong SO coupling and anisotropic g-factor \cite{Winkler2000},  makes 2DHSs promising candidates for quantum information processing with long coherence times \cite{Brunner2009, Marton2023}. In sufficiently clean 2DHSs, the mean-free-paths can be quite long, enabling ballistic transport \cite{Heremans1992, Pfeiffer2005}. Combining ballistic transport with strong and tunable SO coupling can result in unique spin-dependent transport phenomena, with applications in spintronics \cite{Lu1998, Yau2002, Rokhinson2004, Habib2009, Nichele2015, Rendell2022, Rendell2023}.

It is important to reemphasize that these observations have been enabled by decades of innovations in MBE growth techniques \cite{Pfeiffer2003, Manfra2014}. A recent example of MBE innovation is the breakthrough in mobility of GaAs 2D carrier systems following refinements in MBE growth chamber design, and purification of source materials \cite{Chung2021elec,Chung2022elec, Chung2022holes, Chung2018}. Exciting physics shortly followed these mobility breakthroughs, for instance, the appearance of new even-denominator FQHSs in $N=0$ LL of record-quality 2DHSs \cite{Wang2022}. The richness of 2DHSs and recent observations incentivise efforts to further enhance the 2DHS mobility. Given the already extreme levels of vacuum and source material purity in our MBE growth chamber, we present here an alternative approach to improve mobility by optimizing the sample structure design. By systematically growing 60 GaAs 2D hole samples, we find that optimizing two structural parameters, the alloy fraction $x$ of the Al$_x$Ga$_{1-x}$As barriers near the GaAs quantum well (QW), and the QW width $w$, is crucial for maximizing the mobility of 2DHSs. By adjusting these parameters, we obtain significant enhancements in mobility over a wide density range, with a new record value $\simeq$ 1 $\times$ 10$^7$ cm$^2$/Vs, measured at temperature $T=$ 300 mK (Fig.~\ref{fig:fig1}(a)). The improvement achieved at low densities is remarkable given that the previous record $\mu\simeq$ 6 $\times$ 10$^6$ cm$^2$/Vs was achieved at relatively higher density \cite{Chung2022holes}. Interestingly, we also find that at low densities, our 2DHSs display a strong enhancement in mobility as $T$ is lowered from 300 mK to 30 mK, with a record value $\mu\simeq$ 18 $\times$ 10$^6$ cm$^2$/Vs at $p\simeq$ 0.38 $\times$ 10$^{11}$  /cm$^{2}$. Mobilities $>$10$^7$ cm$^2$/Vs are the highest ever achieved for any 2DHS and are bound to unveil new interaction phenomena.

\section{Experimental Methods}

Our samples are grown on 2-inch-diameter GaAs (001) substrates at a growth temperature $T\simeq$ 640 \textdegree C. The ultrahigh vacuum in our growth chamber is achieved by four large (3000 l/s) cryopumps augmented by three auxiliary cryo-cooled ($\simeq$ 17 K) cold plates \cite{Chung2021elec}. The deposition rate is calibrated using refection high-energy electron diffraction oscillations by tuning the oven temperatures. Since our samples use various barrier alloy fractions in the same structure (Fig.~\ref{fig:fig1}(b)), we use two Ga and two Al ovens during each growth and tune the temperatures to obtain the desired alloy fractions. Carbon doping is performed using a doping well structure comprised of a 1.7 nm GaAs QW flanked by 1.13 nm AlAs barriers \cite{ChungDWS}. However, in contrast to 2DESs \cite{Chung2021elec}, the doping well structure does not provide a significant advantage over standard $\delta$-doping into AlGaAs barriers. Carbon doping is achieved using a filament of vitreous C generating a doping rate of $\simeq $10$^{10}$ carbon atoms  /cm$^{2}$s when heated through 6-mm-diameter Ta leads with a power of $\simeq$ 200 W. Typically, the samples are doped for 5 to 10 minutes (depending on the density) by opening a shutter to introduce C atoms and the substrate temperature is reduced to $\lesssim$ 500 \textdegree C. After the doping is completed, the C filament is turned down to a low power ($<$ 1 W) during the growth of undoped regions. 

 On the flanks of GaAs QW, instead of using undoped Al$_x$Ga$_{1-x}$As barriers with constant $x$, we employ stepped barriers with varying $x$ and thicknesses to reduce $x$ near the QW (Fig.~\ref{fig:fig1}(b)). Within this framework, we compare three designs such that $x$ near the QW is 0.32, 0.08 or 0.04; we label the designs by the $x$ value near the QW. Figures~\ref{fig:fig1}(b) and (c) show schematic valence band diagram and a typical MBE structure, respectively, for the $x=$ 0.04 design. The thickness ($s_i$) of each spacer barrier layer is carefully chosen such that no parallel channel forms at any of the step interfaces. The density is tuned by varying the total spacer thickness $s$. The well width is fixed at $w=$ 20 nm for most of the samples, but augmented to $w=$ 30 nm for a study of well-width dependence for $x=$ 0.04.

We characterize the transport properties on unpatterned pieces of the grown wafer, typically 4$\times$4 mm$^2$ in size in the van der Pauw geometry, using standard low-frequency lock-in techniques. The 2DHS is contacted using In:Zn contacts annealed at 450 \textdegree C for 4 minutes in a reducing forming gas  environment. We perform the mobility measurements in dark (without illumination) in a $^3$He cryostat with a base temperature of $\simeq$ 300 mK. Magnetoresistance measurements are then performed to deduce 2DHS density from quantum Hall features. The data presented in Figs.~\ref{fig:fig3} and \ref{fig:fig4} are measured in a dilution refrigerator with a base temperature of $\simeq$ 30 mK.

 \section{Results and Discussion}

Figure~\ref{fig:fig1}(a) shows mobility as a function of 2DHS density for various values of $x$ (closest to the QW) and $w$. Focusing on the $w=20$ nm data, for $p\leq$ 1 $\times$ 10$^{11}$ /cm$^{2}$, a factor of $\simeq$ 3 improvement is seen as $x$ is lowered from 0.32 to 0.08, and another factor of $\simeq$ 1.5 when lowered to 0.04. We conjecture that this dramatic improvement stems from two effects. First, the purity of the Al source even after sufficient cleaning is not as high as the Ga source \cite{Chung2018, Chung2021elec}. Additionally, Al atoms are much more chemically reactive than Ga atoms and can attract stray impurities from the imperfect vacuum environment more readily and incorporate them into the structure during growth. Using lower $x$ reduces the concentration of background impurities near and in the QW \cite{Chung2018, Chung2021elec}. Second, lower $x$ also leads to a smoother and more gradual potential profile at the barrier-QW interface. This gentle confinement of holes in the QW reduces sensitivity to small variations or imperfections at the interface, lowering the interface roughness. Additional evidence for reduction in interface roughness comes from the fact that when $w$ is increased to 30 nm for $x=$ 0.04, mobility improves further. Lowering $x$ or increasing $w$ further does not show more improvement, likely because of weaker confinement of the 2DHS in the GaAs QW \cite{Supp}.

In Fig.~\ref{fig:fig1}(a), at high densities, mobility falls rapidly with increasing $p$, with $x=$ 0.04 data falling faster than either 0.08 or 0.32. At higher $p$, the penetration of the wavefunction into the barrier can lead to alloy-disorder scattering. The penetration increases with decreasing $x$ because of the lower potential barrier. The contribution of remote ionized impurities (the intentional impurities in the doping region), also increases at higher $p$ as the setback $s$ becomes smaller. This brings remote ionized impurities closer to the QW, increasingly degrading the mobility. 

We now discuss in detail the various disorder effects that can account for Fig.~\ref{fig:fig1}(a) results. Qualitatively, we have identified that residual \textit{background impurities} (BIs) in the channel and barrier, \textit{interface roughness} (IR), \textit{remote ionized impurities} (RIs), and \textit{alloy disorder} (AD) are the main factors that limit the mobility in our 2DHSs. We analyze these scattering mechanisms quantitatively using transport models and obtain the dependence on $p$ for each scattering mechanism. In the simple Drude picture, the low-temperature mobility is defined as $\mu= e\tau/m^*$ where $e$ is the fundamental charge and $\tau$ is the total scattering lifetime. For each scattering mechanism, it is useful to define a characteristic lifetime $\tau_j$ where the subscript $j$ marks the scattering mechanism under consideration. Using Mathiessen’s rule, $\tau$ can then be evaluated as $\frac{1}{\tau}=\sum\limits_{j} \frac{1}{\tau_j}=\frac{1}{\tau_\textrm{BI}}+\frac{1}{\tau_\textrm{IR}}+\frac{1}{\tau_\textrm{RI}}+\frac{1}{\tau_\textrm{AD}}$.

An important characteristic of 2DHSs is that $m^*$ has a strong dependence on $p$ (Fig.~\ref{fig:fig1}(d)) arising from non-parabolicity of the valence band due to mixing of the heavy-hole and light-hole bands \cite{Winkler2003, Zhu2007, Watson2012, Chung2022holes}. To take this into account, we calculate hole energy-band dispersions for each case in Fig.~\ref{fig:fig1}(a) and then determine the density-of-states (DOS) effective mass at the Fermi energy (Fig.~\ref{fig:fig1}(d)) which we use as $m^*$ in our transport models. Our self-consistent calculations are based on an 8$\times$8 Kane Hamiltonian with cubic anisotropy but no Dresselhaus term \cite{Winkler2003,footnote1}. A dependence of $m^*$ on $x$ and $w$ is also evident in Fig.~\ref{fig:fig1}(d).  The decline in mobility at higher $p$ in Fig.~\ref{fig:fig1}(a) appears to be correlated with the rise in $m^*$ in Fig.~\ref{fig:fig1}(d) which is expected from the Drude picture. More importantly, this strongly density-dependent $m^*$ can significantly affect the density dependence of scattering mechanisms and needs to be carefully incorporated into the transport models. 

Following the Born approximation, the general form of $\tau_j$ can be written as \cite{Gold1989, Huang2022}:
\begin{equation}
\frac{1}{\tau_j} = \frac{m^*}{2\pi\hbar^3k_F^3}\int_0^{2k_F}dq\frac{q^2}{\sqrt{1-(\frac{q}{2k_F})^2}}\frac{\braket{|U_j(q)|^2}}{\epsilon_q^2},
\label{eq:tau}
\end{equation}  
where integration is over the wavevector $q$, $k_F=\sqrt{2\pi p}$ is the Fermi wavevector, $\hbar$ is Planck's constant, $U_j(q)$ is the scattering potential for a given scattering mechanism, and $\epsilon_q$ is the dielectric screening function which, under the random phase approximation, is given by:
\begin{equation}
\epsilon_q = 1 + \frac{q_s}{q}F_c(q)[1-G(q)].
\label{eq:epsilon}
\end{equation}  
Here $q_s$ is the screening wavevector which becomes $q_\textrm{TF}=\frac{m^*e^2}{2\pi\epsilon_b\epsilon_0\hbar^2}=\frac{2}{a^*_B}$ in the Thomas-Fermi approximation, $\epsilon_0$
the vacuum permittivity, $\epsilon_b=$ 12.9 the dielectric constant of GaAs and $a^*_B$ the effective Bohr radius. The interaction effects in the 2DHS are accounted by a local-field correction term $G(q)=\frac{q}{2\sqrt{q^2+k_F^2}}$ within the Hubbard approximation, and a form factor $F_c(q)$ for hole-hole Coulomb interaction to take into account the finite confinement, given by \cite{Gold1989, Huang2022}:
\begin{equation}
F_c(q)=\int_{-\infty}^{+\infty}dz|\psi(z)|^2\int_{-\infty}^{+\infty}dz'|\psi(z')|^2e^{-q|z-z'|}.
\label{eq:ff}
\end{equation}
We note that exchange-correlation effects can be more complex in 2DHSs as compared to 2DESs, for instance because of an anomalous-spin polarization in 2D holes \cite{Winkler2005}. Also, the hole wavefunctions are four-component spinors (representing the effective spin 3/2 of holes) for finite in-plane wavevector $k_{||}$ \cite{Winkler2003}. For simplicity, we evaluate the wavefunction $\psi(z)$  self-consistently at the subband edge ($k_{||}=$ 0) for each $x$ and $w$ in Fig.~\ref{fig:fig1}(a) at a given $p$. We now discuss each scattering mechanism separately and evaluate $\braket{|U_j(q)|^2}$.

\textit{Residual background impurities, BIs}. Even if the MBE chamber is ultraclean, it is not completely devoid of impurities, and these could get incorporated in the structure during growth. If charged, they can scatter holes. The average potential for BIs is evaluated using Eqs.~\ref{eq:BI} and \ref{eq:ff2}. It is important to distinguish between BIs according to where they reside, in the GaAs QW or in the Al$_x$Ga$_{1-x}$As barrier, because concentration of BIs in the barrier ($N_{\textrm{BI.}s_i}$) can be significantly higher than in the QW ($N_{\textrm{BI.}w}$). Accordingly, we define BI density $N(z)$ as a function of distance $z$ in Eq.~\ref{eq:NBI}. The highest-mobility samples with $x=0.04$ design set the level of BIs, and we find $N_{\textrm{BI.}w}=$ 2 $\times$ 10$^{12}$ /cm$^{3}$ and $N_{\textrm{BI.}s_1}=$ 5 $\times$ 10$^{12}$ /cm$^{3}$. Considering higher concentration of BIs in the barriers with higher $x$ and concomitant surface segregation into the QWs \cite{Chung2018}, we find $N_{\textrm{BI.}w}$ and $N_{\textrm{BI.}s_i}$ for other designs (see Supplemental information for more details \cite{Supp}).

\begin{figure}[!t]
\begin{center}
\includegraphics[width=0.5 \textwidth]{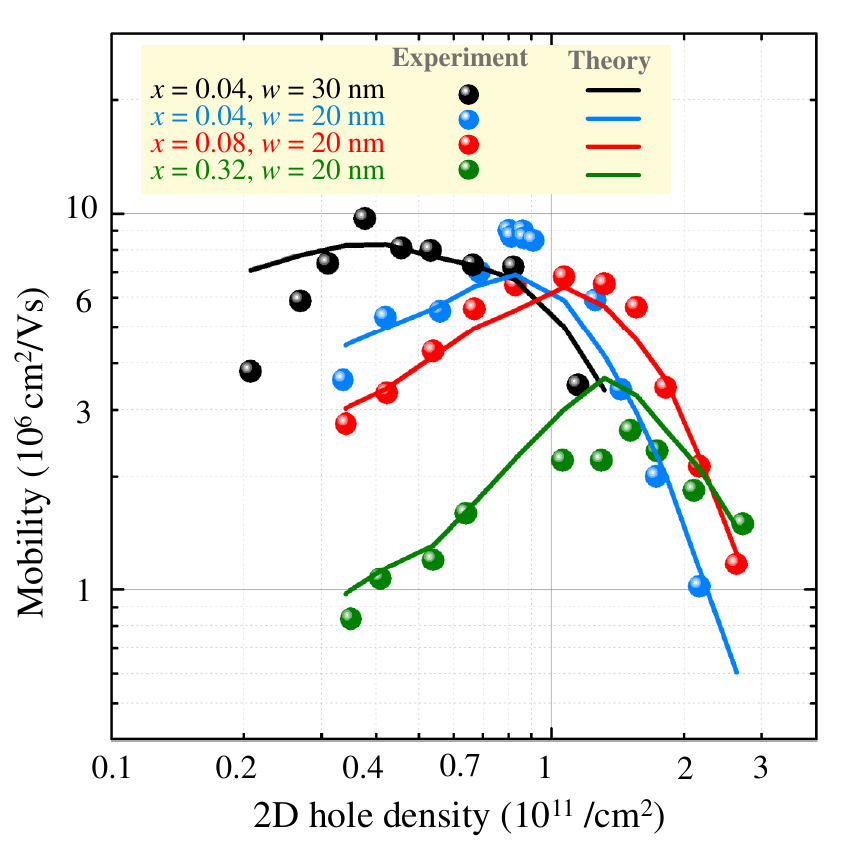}
\caption{ Measured 2DHS mobility vs density in comparison with calculated mobility limited by the combined scattering mechanisms. The contributions of individual scattering mechanisms are shown in Fig. S2 \cite{Supp}. \label{fig:fig2}}
\end{center}
\end{figure}

\textit{Interface Roughness, IR}. Scattering from IR results from layer variations at the GaAs/AlGaAs interface which cause fluctuations in QW width, ground-state energy, and local charge distribution, creating a scattering landscape for holes. We employ a model for finite QWs where the well width fluctuations are parameterized by two parameters, the average height of fluctuations and the correlation length over which the fluctuation spreads. The averaged random potential takes the form in Eq.~\ref{eq:IR}. It is worth noting that the density dependence of IR in our 2DHSs (positive slope) is in contrast to 2DESs (negative slope) reported recently \cite{Chung2022elec}. This is because IR is strongly dependent on $w$, and $w$ is fixed here for the entire range of density in 2DHSs while it was decreased with density in 2DESs \cite{Chung2022elec} (causing more scattering at higher densities). A combination of BI and IR gives a reasonable agreement with the experimental data at low densities \cite{Supp}.

\begin{figure*}[!t]
\begin{center}
\includegraphics[width=1 \textwidth]{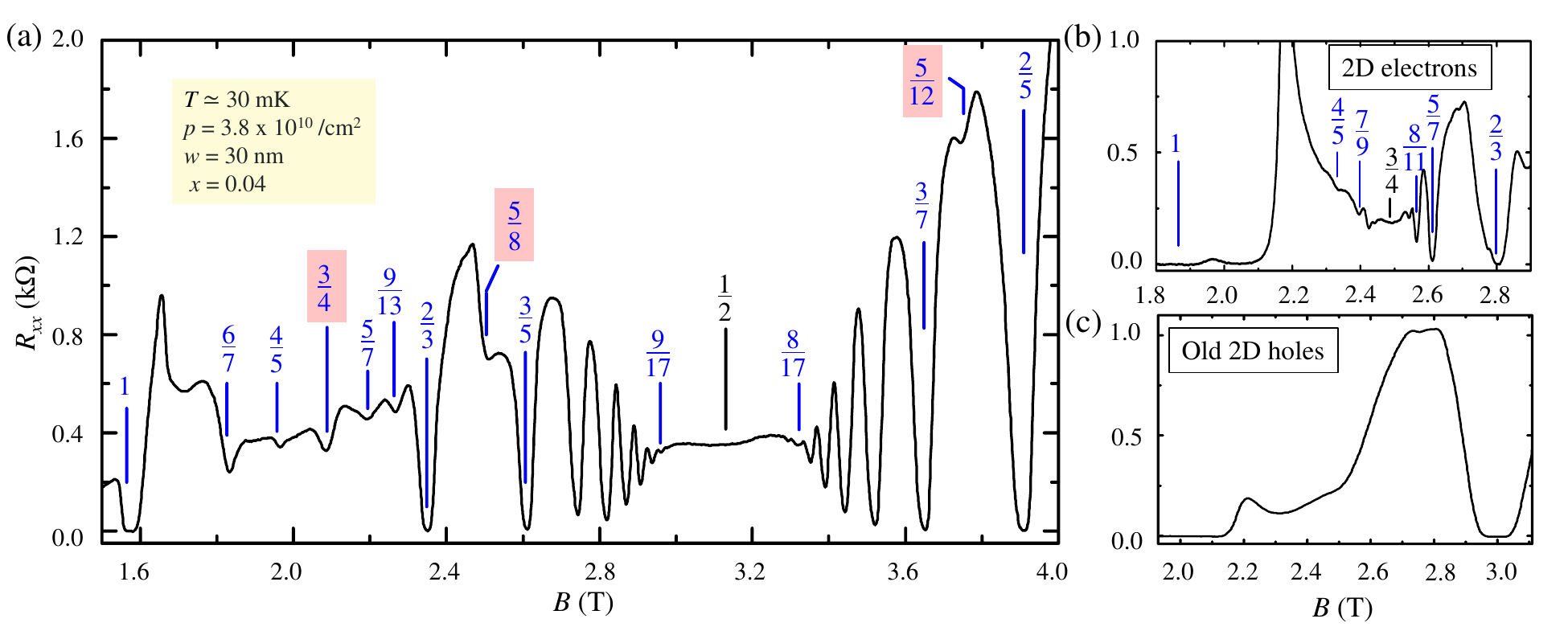}
\caption{(a) Longitudinal resistance ($R_{xx}$) vs perpendicular magnetic field $B$ of our record-high-mobility sample at $p=$ 0.38$\times10^{11}$  /cm$^{2}$ with $x=$ 0.04 and $w=$ 30 nm at $T\simeq$ 30 mK. The magnetic field positions of several QHSs are marked in blue. The developing even-denominator FQHSs are highlighted in red. (b) $R_{xx}$ vs $B$ of a high-mobility ($\mu\simeq10\times10^6$ cm$^2$/Vs at 300 mK) 2D electron sample, with a density of $\simeq0.45\times10^{11}$, $w=40$ nm, at $T\simeq$ 30 mK. (c) $R_{xx}$ vs $B$ of an old 2D hole sample ($\mu\simeq1\times10^6$ cm$^2$/Vs at 300 mK) with $p\simeq0.48\times10^{11}$, $x=0.32$, and $w=30$ nm, at $T\simeq$ 40 mK. The magnetoresistance traces in (b) and (c) are shown in a narrow $B$ range between $\nu=1$ and $\nu=2/3$ to highlight the differences from (a). \label{fig:fig3}}
\end{center}
\end{figure*}

\textit{Remote ionized impurities, RIs}. RI scattering is inevitable in a modulation-doped structure and comes from the long-range Coulomb interaction between the holes in the QW and their parent ions in the doping layers. In our calculations, we assume that the sheet density of RIs ($n_\textrm{RI}$) is equal to the 2DHS density in the QW, thus neglecting surface compensation contribution to $n_\textrm{RI}$. Similar to BIs, the random potential for RIs takes the form shown in Eq.~\ref{eq:RI}. While RI scattering shows the right trend vs density, it proves insufficient to explain the sharp decline in $\mu$ at higher densities.

\textit{Alloy disorder, AD}. In the  Al$_x$Ga$_{1-x}$As barriers, the Al and Ga atoms are randomly distributed, which can cause localized potential fluctuations. The tails of the wavefunctions extending into the barrier can interact with these random potential fluctuations and cause scattering. Using virtual crystal approximation, the AD potential can be written as Eq.~\ref{eq:AD}. As expected, AD affects $x=0.04$ the most because of the maximum penetration into the barrier (see Supplemental material Figs. S2 and S4), and as a result mobility falls much faster for $x=0.04$ at higher $p$.

Combining the contributions of the above scattering mechanisms, the resultant total mobility $\mu=$ e$\tau$/$m^*$ for each design, plotted in Fig.~\ref{fig:fig2}, is in reasonably good agreement with the experimental data. Our calculations capture the salient features of the experimental data - the non-monotonic dependence of mobility on density, and the crossings between mobilities of different designs at higher densities.

We acknowledge of course, the possibility of other scattering mechanisms such as intersubband scattering between electric subbands or between spin-split subbands. For narrow QWs ($w=20$ nm), our self-consistent calculations suggest that the second electric subband is not occupied throughout the density range considered. For wide wells ($w=30$ nm, $x=$ 0.04), there is a possibility of second subband occupation for $p\geq$ 1.5$\times10^{11}$  /cm$^{2}$, but we stay below that range in experiments. In principle, inter-spin-split-subband scattering in the presence of SO coupling can become relevant in 2DHSs \cite{Hwang2003}. However, under relevant measurement circumstances, this scattering is usually weak \cite{footnote2} and thus ignored in the transport calculations. Another mechanism which could become pertinent at very low densities is the density inhomogenity induced percolation. Indeed, our mobilities exhibit a faster decay at very low densities  ($p\lesssim$ 0.4$\times10^{11}$) which can be explained using percolation models \cite{Sojoto1990,Manfra2007}. Fitting our $w=30$ nm data for conductivity $\sigma$ at very low densities to $\sigma \sim (p - p_c)^{\delta}$, where $p_c$ is the critical percolation density and $\delta$ is the critical exponent \cite{Sojoto1990,Manfra2007}, we find reasonable values of $p_c\simeq$ 7$\times$ 10$^9$ /cm$^{2}$ and $\delta\simeq1.9$, suggesting mobilities at $p\lesssim3\times$ 10$^{10}$ /cm$^{2}$ may be incipiently affected by density inhomogeneities.

As discussed earlier, one of the most important applications of high-mobility 2D carrier systems lies in probing many-body phenomena. Given the remarkable improvement in mobility, we studied the low-$T$ ($\simeq$ 30 mK) magnetotransport characteristics in our record high-mobility 2DHS at  $p=$ 3.8$\times10^{10}$ /cm$^{2}$ ($x=$ 0.04 and $w=$ 30 nm). Figure~\ref{fig:fig3}(a) shows $R_{xx}$ vs perpendicular magnetic field $B$, with several LL fillings marked. Clearly, the sample exhibits exceptional quality as evinced by numerous even- and odd-denominator FQHSs. Along with the even-denominator $\nu=$ 3/4 FQHS recently observed in ultraclean 2DHSs \cite{Wang2022}, several other delicate features are observed between $\nu=$ 1 and 2/3 at $\nu=$ 6/7, 4/5, 5/7 and 9/13, suggesting developing FQHSs at these fillings. Apart from $\nu=$ 3/4, $R_{xx}$ minima are observed at other even-denominator $\nu=$ 5/8 and 5/12. The trace in Fig.~\ref{fig:fig3} also shows numerous higher-order odd-denominator FQHSs near $\nu=$ 1/2 up to  $\nu=$ 9/19. Such higher-order FQHSs in a very low-density sample again attest to the quality of our 2DHS. In the Supplemental Figs.~ S3 and S4, we show more examples of traces taken at $T=$ 300 mK to corroborate that higher mobility samples indeed show more and better-defined FQHSs.

In Figs. \ref{fig:fig3}(b) and (c), we compare magnetotranport data of Fig. \ref{fig:fig3}(a) with previously grown 2DES and 2DHS samples, respectively, between $\nu=1$ and 2/3, at a comparable density, QW width and temperature. The 2DES shows a reenterant integer QHS at $B\simeq2.1$ T,  and standard (Jain-sequence) odd-denominator FQHSs at $\nu=$ 4/5, 7/9 and 5/7, 8/11,... flanking a smooth $R_{xx}$ minimum at $\nu=$ 3/4. Clearly, these features are very different from our dilute high-mobility 2DHS (Fig. \ref{fig:fig3}(a)) which exhibits additional developing FQHSs at  $\nu=$ 6/7 and 9/13, and an even-denominator developing FQHS at $\nu=$ 3/4. The 2DHS in Fig. \ref{fig:fig3}(c) which has much lower mobility, shows essentially no FQHSs between $\nu=1$ and 2/3.

\begin{figure}[!t]
\begin{center}
\includegraphics[width=0.5 \textwidth]{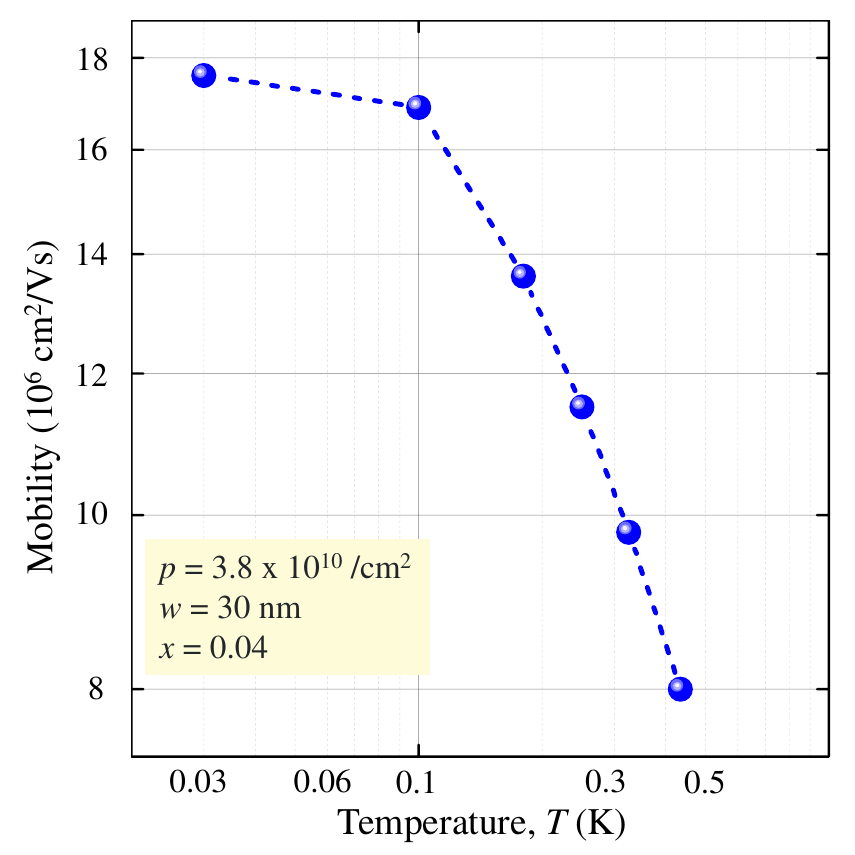}
\caption{Measured mobility vs temperature in the sample with $x=0.04$, $w=$ 30 nm, and $p\simeq$ 0.38$\times10^{11}$  /cm$^{2}$. The temperature is varied from 30 mK to 300 mK. The enhancement of $\mu$ at lower $T$ indicates a metallic behavior. \label{fig:fig4}}
\end{center}
\end{figure}

We finally discuss a rather dramatic enhancement in mobility of our record-high-mobility, dilute 2DHS as we lower the temperature (Fig.~\ref{fig:fig4}). In Fig.~\ref{fig:fig2}, the highest mobility at $T=$ 300 mK is $\simeq$ 10 $\times$ 10$^6$ cm$^2$/Vs at $p\simeq$ 0.38$\times10^{11}$ /cm$^{2}$ ($w=30$ nm, $x=$ 0.04). As we cool this sample, the mobility increases to $\simeq18\times$ 10$^6$ cm$^2$/Vs at $T\simeq$ 30 mK (an improvement by a factor of $\simeq$ 1.8). We note that similar behaviors have been observed in other dilute GaAs 2DHSs \cite{Hanein1998, Simmons1998, Watson2011}. In fact, at similar densities, Watson \textit{et al.} \cite{Watson2011} also report an increase in mobility by a factor $\simeq1.8$ when $T$ is lowered from 300 mK to 50 mK in their 2DHSs. This mobility increase in dilute carrier systems can stem from the temperature dependence of dielectric screening  \cite{ Sarma2015}. It is more pronounced in GaAs 2DHSs as compared to 2DESs because of the larger $m^*$, which enhances the dimensionless parameter $q_\textrm{TF}/2k_F$ ($\simeq$13 for the sample in Fig.~\ref{fig:fig4}), making screening more effective.

While screening provides a possible explanation for the observed temperature dependence, for our sample parameters ($m^*$, $p$ and $T_F$), a factor of $\simeq$ 1.8 is too high to be attributed entirely to screening \cite{Sarma2015,footnote3}. We surmise that some other, yet unknown factors may be contributing to the temperature dependence in dilute 2DHSs, apart from screening \cite{footnote3}. More careful temperature-dependent mobility measurements at different densities may help illuminate the underlying mechanism. The temperature dependence of mobility can also affect the density dependence of mobility in Fig.~\ref{fig:fig1}(a), particularly at lower densities. Since background impurities mostly limit the mobility at lower densities (in particular for the case depicted in Fig.~\ref{fig:fig4} \cite{Supp}), that would imply our MBE-grown GaAs has an even lower concentration of background impurities ($N_{\textrm{BI.}w} \lesssim 1\times10^{12}$ /cm$^{3}$) than we estimated from Fig.~\ref{fig:fig2} fits.

\section*{Acknowledgements}
We acknowledge support by the National Science Foundation (NSF) Grants Nos. DMR 2104771 and ECCS 1906253) for measurements, the U.S. Department of Energy (DOE) Basic Energy Sciences Grant No. DEFG02-00-ER45841 for sample characterization, and the Eric and Wendy Schmidt Transformative Technology Fund and the Gordon and Betty Moore Foundation’s EPiQS Initiative (Grant No. GBMF9615 to L.N.P.) for sample fabrication. Our measurements were partly performed at the National High Magnetic Field Laboratory (NHMFL), which is supported by the NSF Cooperative Agreement No. DMR 2128556, by the State of Florida, and by the DOE. This research was funded in part by QuantEmX travel grant from Institute for Complex Adaptive Matter (ICAM) and the Gordon and Betty Moore Foundation through Grant GBMF9616 to A. G., C. W., S. K. S., and M. S.

\appendix
\section*{Appendix: Details of transport calculations}
\setcounter{equation}{0}
\renewcommand{\theequation}{A\arabic{equation}}
In order to calculate $\tau_j$ for a given scattering source, we need to calculate the square of averaged random potential $\braket{|U_j(q)|^2}$.

\textit{Background impurities (BIs)}. The potential for BIs averaged over impurity positions can be written as \cite{Gold1989, Huang2022}:
\begin{equation}
\braket{|U_{\textrm{BI}}(q)|^2}=\left(\frac{e^2}{2\epsilon_b\epsilon_0q}\right)^2\int_{-\infty}^{+\infty}dz N(z)F_{imp}^2(q,z).
\label{eq:BI}
\end{equation}
Here, $N(z)$ is the three-dimensional impurity concentration at a distance $z$ from the center of the QW, and  $F_{imp}(q,z)$ is the form factor for hole-impurity interaction which takes into account the finite width of the QW and is given by:
\begin{equation}
F_{imp}(q,z)=\int_{-\infty}^{+\infty}dz'|\psi(z')|^2e^{-q|z-z'|}.
\label{eq:ff2}
\end{equation} 
We note that for the barrier layers, BIs in the layers closest to the QW contribute the most to limit the mobility, and the contribution decreases exponentially for the subsequent layers. We take into account BIs in the first two Al$_x$Ga$_{1-x}$As layers (for example, $x=0.04$ and $x=0.06$ with thicknesses $s_1$ and $s_2$, respectively, in Fig.~\ref{fig:fig1}(b)). Assuming a homogeneous distribution of BIs, $N(z)$ can be defined as:

\begin{equation}
\
N(z) =
\begin{cases}
    N_{\textrm{BI.}w} &  |z| < w/2, \\
    N_{\textrm{BI.}s_1} &  w/2 \leq |z| < w/2+s_1, \\
    N_{\textrm{BI.}s_2}&  w/2+s_1 \leq |z| < w/2+s_1+s_2.
\end{cases}
\
\label{eq:NBI}
\end{equation}

\textit{Interface roughness (IR)}. The IR is characterized using an autocorrelation function $\braket{\Delta(r)\Delta(r')}=\Delta^2$exp($-|r-r'|^2/\Lambda^2$) which defines the correlation between fluctuations at different points along the interface;  $\Delta$ is the height of fluctuations and $\Lambda$ is the lateral size. The random potential for a finite barrier takes the form \cite{Li2005}:
\begin{equation}
\braket{|U_\textrm{IR}(q)|^2}=\pi\Lambda^2\Delta^2F_\textrm{IR}^2e^{-\frac{\Lambda^2q^2}{4}},
\label{eq:IR}
\end{equation}
where $F_\textrm{IR}$ is the function characterizing the local change in ground state energy $E_0$ with respect to change in $w$, such that:
\begin{equation}
F_\textrm{IR}=\frac{\partial E_0}{\partial w}= -\frac{2E_0}{\sqrt{\frac{2\hbar^2}{m^*(V-E_0)}}+w}.
\label{eq:FIR}
\end{equation}
We also assume that both interfaces contribute equally to total IR with same roughness parameters. We find $\Delta \lesssim$ 3.4 \AA~ and $\Lambda \simeq 25$ nm fits the data well for all the designs .

\textit{Remote ionized impurities (RIs)}. Confined to a 2D plane at a distance s from the 2DHS, RIs lead to a random potential of the form:
\begin{equation}
\braket{|U_{\textrm{RI}}(q)|^2}=\left(\frac{e^2}{2\epsilon_b\epsilon_0q}\right)^2n_\textrm{RI}F_{imp}^2(q,s),
\label{eq:RI}
\end{equation}
where $F_{imp}(q,s)$ is same as the form factor in Eq.~\ref{eq:ff2}, evaluated at $z=s$. In our calculations, we take the distance from the center of the QW such that $z=s+w/2$.

\textit{Alloy disorder (AD)}. The random averaged potential for AD can be written using the virtual crystal approximation, as \cite{Ando1982}:
\begin{equation}
\braket{|U_\textrm{AD}(q)|^2}=V_\textrm{AD}^2\Omega x(1-x)F_\textrm{AD},
\label{eq:AD}
\end{equation}
 where $V_\textrm{AD}$ characterizes the strength of AD \cite{Gold1989}; $\Omega= \frac{a^3}{4}$ is the volume element of the alloy unit cell with $a=$ 5.67~\AA~ the lattice constant in GaAs, and $F_\textrm{AD}$ is the form factor for AD given by \cite{Ando1982}:
\begin{equation}
F_\textrm{AD}=\int_{\textrm{barrier}}dz|\psi(z)|^4.
\label{eq:ffAD}
\end{equation}
From fits to our data, we find that the parameter $V_\textrm{AD}$ = 0.65, 0.75, and 3.2 eV for $x=$ 0.04, 0.08 and 0.32, respectively \cite{Supp}.



\section*{Supplementary material}

Supplemental Material to

# Ultraclean two-dimensional hole systems with mobilities exceeding $10^7$ cm$^2$/Vs

Adbhut Gupta,[1] C. Wang,[1] S.K. Singh,[1] K.W. Baldwin,[1] R. Winkler,[2] M. Shayegan,[1] and L.N. Pfeiffer[1]

[1]*Department of Electrical and Computer Engineering, Princeton University, Princeton, NJ, USA*

[2]*Department of Physics, Northern Illinois University, DeKalb, Illinois, USA*

# S1 Attempts to further improve mobility

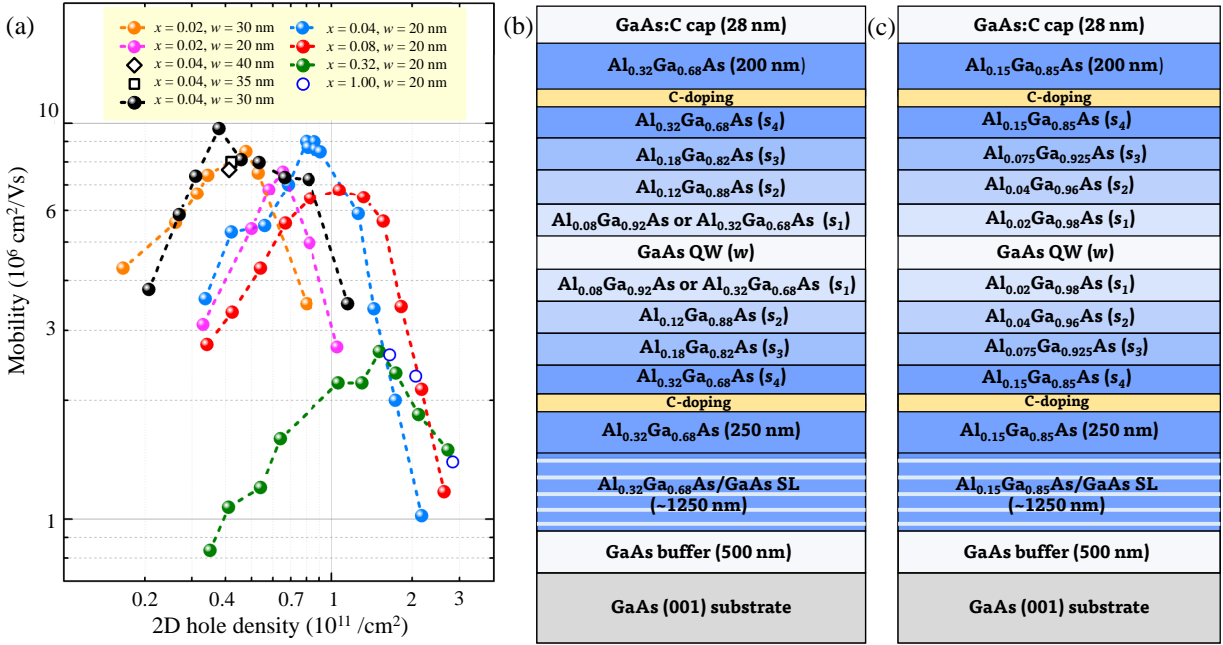

Fig. S1. (a) Extension of Fig. 1(a) of the main text with Mobility vs 2D hole density data for all 60 samples included. (b) Layer structures for the $x = 0.08$ and $x = 0.32$ designs. The only difference is the Al composition of barrier layers closest to the QW. (c) Layer structure for the $x = 0.02$ design.

Following improvements in mobility as discussed in the main text, a straightforward approach for further improvement is to lower the Al content near the quantum well (QW) to even smaller values. Accordingly, we grew samples with $x = 0.02$ near the QW as shown in Fig. S1(a). For $w = 20$ nm (pink circles in Fig. S1(a)), there is no significant improvement compared to the $x = 0.04$ design for $p \leq 0.7 \times 10^{11}$ /cm$^2$. Increasing the density further, in fact, results in a dramatic fall of mobility because of alloy disorder scattering becoming relevant at relatively lower densities for $x = 0.02$ (more severe wavefunction penetration into the barrier). Increasing $w$ to 30 nm (orange circles in Fig. S1(a)) enhances the mobility compared to 20 nm, but again not much improvement is observed compared to $x = 0.04$, $w = 30$ nm.

Next we test the effect of increasing QW width in $x = 0.04$ design while keeping density fixed at $\simeq 0.4 \times 10^{11}$ /cm$^2$. In the main text, we already discussed the improvement in mobility (a factor of $\simeq 2$) as $w$ is increased from 20 nm to 30 nm. We increase $w$ further to 35 nm and 40 nm. We measure $\mu = 8.1 \times 10^6$ ($7.6 \times 10^6$) cm$^2$/Vs for $w = 35$ (40) nm. Clearly, $w = 30$ nm with $\mu = 9.7 \times 10^6$ cm$^2$/Vs is the optimum QW width at $p \simeq 0.4 \times 10^{11}$ /cm$^2$.

| $x$ ($w$ nm) | $p$ ($10^{11}$ /cm$^2$) | $\mu$ ($10^6$ cm$^2$/Vs) | $s_1$ (nm) | $s_2$ (nm) | $s_3$ (nm) | $s_4$ (nm) |
|---|---|---|---|---|---|---|
| 0.04 (30) | 0.21 | 3.8 | 80 | 80 | 160 | 400 |
| | 0.38 | 9.7 | 40 | 40 | 80 | 200 |
| | 0.66 | 7.3 | 25 | 25 | 50 | 125 |
| | 1.15 | 3.5 | 15 | 15 | 30 | 75 |
| 0.04 (20) | 0.41 | 5.3 | 40 | 40 | 80 | 200 |
| | 0.83 | 9 | 20 | 20 | 40 | 100 |
| | 1.26 | 5.9 | 13 | 13 | 26 | 56 |
| | 2.17 | 1 | 8 | 8 | 16 | 40 |
| 0.08 (20) | 0.54 | 4.3 | 50.8 | 76.2 | 101.5 | 254 |
| | 1.07 | 6.8 | 26 | 39 | 52 | 130 |
| | 1.56 | 5.6 | 17.7 | 26.5 | 35.4 | 88.4 |
| | 2.17 | 2.1 | 12 | 18 | 24 | 60 |
| 0.32 (20) | 0.41 | 1.1 | 63.5 | 95.2 | 127 | 317.4 |
| | 1.06 | 2.2 | 26 | 39 | 52 | 130 |
| | 1.51 | 2.7 | 17.7 | 26.5 | 35.4 | 88.4 |
| | 2.72 | 1.5 | 9.6 | 14.4 | 19.3 | 48.1 |

Table SI. Structure parameters from some representative 2DHSs in various density ranges in our study. All mobility values listed are from measurements at 0.3 K.

As discussed in the main text, at higher densities, alloy disorder scattering limits the mobility because of larger wavefunction penetration into the barrier. To restrict the wavefunction in the QW, we modify the $x = 0.04$, $w = 20$ nm structure by adding thin layers (1.13 nm) of pure AlAs ($x = 1$) near the QW. Indeed, compared to the $x = 0.04$ structure (solid blue circles in Fig. S1(a)), the mobility shows a significant improvement for $p \gtrsim 1.7 \times 10^{11}$ /cm$^2$ (open blue circles). However, pure AlAs increases the background impurities locally near the QW. While we are able to reduce alloy disorder scattering, this competing mechanism prevents the mobility to show significant improvement over the $x = 0.32$ design.

Figures S1(b),(c) show typical layer structures for the $x = 0.08$, 0.32 and 0.02 designs. The only difference between the $x = 0.08$ and $x = 0.32$ designs is the Al composition of the barrier layer closest to the QW. As discussed in the next section, simply changing one layer can affect the mobility significantly by modifying the level of background impurities. In Table S1, we show the structural parameters for selected samples from Fig. 1(a) main text.

## S2 Scattering mechanisms

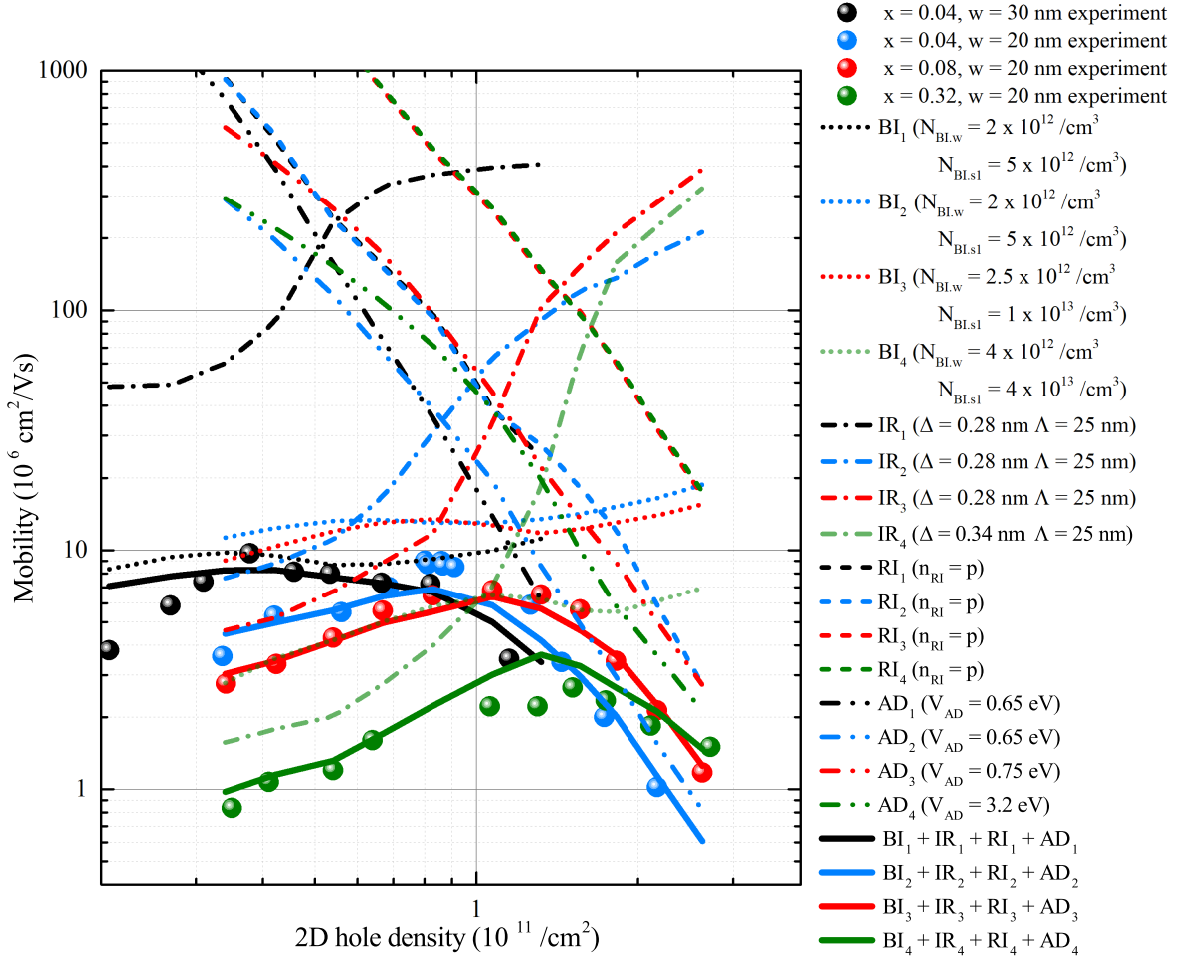

Fig. S2. Mobility vs density for various scattering mechanisms. Solid circles represent the experimental data. The four designs discussed in the main text are: $x = 0.04$, $w = 30$ nm; $x = 0.04$, $w = 20$ nm; $x = 0.08$, $w = 20$ nm; and $x = 0.32$, $w = 20$ nm. Mobilities ensued from scattering by residual background impurities (BIs) are represented by dotted lines, interface roughness (IR) by dash-dotted lines, remote ionized impurities (RIs) by dashed lines, and alloy disorder (AD) by dash-dot-dot lines. The solid lines represent the total mobility obtained by adding individual mechanisms according to Matthiessen's rule. Different colors represent different designs.

In Fig. 2 of the main text, we showed the final calculated mobility vs density overlaid on experimental data. Here we discuss the contributions of individual scattering mechanisms as a function of density. The calculations are done using the models described in the Methods section of the main text. Figure S2 shows the results from different scattering mechanisms plotted along with the experimental data. Starting with low densities ($\leq 1\times 10^{11}$ /cm$^2$), we first include background

4

impurities both in the GaAs QW and the two layers of AlGaAs barriers closest to the QW. Because of the strong dependence of effective mass $m^*$ on density (Fig. 1(d), main text), BI shows a non-monotonic dependence as a function of density (dotted lines). The highest-mobility design ($x = 0.04$, $w = 30$ nm) sets the concentration of BIs and we get $N_{\text{BI}.w} = 2 \times 10^{12}$ /cm$^3$ and $N_{\text{BI}.s_1} = 5 \times 10^{12}$ /cm$^3$. For other designs, we calculate $N_{\text{BI}.w}$ and $N_{\text{BI}.s_1}$ based on two assumptions. First, we assume that $N_{\text{BI}.s_1}$ scales linearly with $x$, for instance, $x = 0.32$ design has 8 times more Al impurities than $x = 0.04$ design. Our second assumtion is that, $N_{\text{BI}.w}$ increases with increasing $x$ because of the surface segregation effect [1] which can locally enhance the BIs in the QW. We assume that 5% of the total impurities in the barriers get dumped in the QW through surface segregation. Accordingly, we get BI concentrations as shown in Fig. S2. Yet, within the constraints discussed above, considering only residual BIs does not explain our low-density data for all the data sets.

Accounting for interface roughness (IR) scattering helps fit the lower-density data better, as shown in Fig. S2 (dash-dotted lines). The parameter $\Lambda = 25$ nm is chosen such that it matches the slope of $\mu$ vs $p$ at low $p$ and parameter $\Delta$ is kept at 1 monolayer thickness (2.83 Å) for most designs. However, for $x = 0.32$ we need to assume slightly higher value $\Delta \simeq 3.4$ Å. While for epitaxial growth IR should exhibit a step-like behavior, where each step is an integral multiple of monolayer thickness, we surmise that the presence of impurities and/or random fluctuations can lead to an irregular roughness profile where the average height of the roughness can deviate slightly from monolayer thickness. It is clear from the $x = 0.04$ design that increasing the QW width makes IR much weaker and then BI becomes the dominant scattering mechanism. Importantly, IR can play a significant role in limiting the mobility at lower densities, in addition to BIs, for narrow QWs. Hence, for maximizing mobility, the QW width must be optimized. With the assumptions described above, we obtain reasonable fits to the experimental data at low densities using a combination of BI and IR.

For higher densities, we start with remote ionized impurities (RIs). Assuming that the density of RIs ($n_{RI}$) is the same as the density of holes ($p$) makes the treatment of RI parameter-free. For $x = 0.08$ and $x = 0.32$, RI curves overlap because their setback distance is the same (Fig. S1(b)). For $x = 0.04$, the setback distance is almost a factor of $\simeq 2$ smaller because the value of $x$ near the doping is 0.17 for the $x = 0.04$ design vs 0.32 for the other two designs (see Figs. 1(b),(c) of main text and Fig. S1(b)). The declining trend of $\mu_{\text{RI}}$ vs $p$ is consistent with a decreasing setback with increasing $p$ but the RI is not strong enough to affect the mobility considerably at higher $p$, particularly for $x = 0.08$ and $x = 0.32$.

Finally, we include the alloy disorder (AD) scattering. The negative slope of $\mu_{\text{AD}}$ vs $p$ seems to

match the slope of the experimental data. Adjusting the parameter $V_{\text{AD}}$ helps fit the experimental data reasonably at high densities. We obtain $V_{\text{AD}} = 0.65$ and $0.75$ eV for $x = 0.04$ and $0.08$ respectively. For $x = 0.32$, we need to assume a much higher $V_{\text{AD}} = 3.2$ eV. While penetration of wavefunction into the barrier is smaller for $x = 0.32$, the amount of disorder in the barrier increases merely because of more Al atoms. Increasing value of $V_{\text{AD}}$ with $x$ indicates that it might relate to the strength of alloy disorder potential [2] (see discussion of Fig. S4). However, because of the uncertainty in the definition [3] and reported values [4], we simply treat $V_{\text{AD}}$ as a free parameter without going into details of its physical significance. We note that the virtual crystal approximation used to treat AD assumes $\text{Al}_x\text{Ga}_{1-x}\text{As}$ to be a homogeneous material with an effective composition, ignoring any short-range order or local distortions around atoms. This goes beyond the definition of alloy disorder and more rigorous theories may be needed to model it.

## S3  Magentoresistance traces at 0.3 K

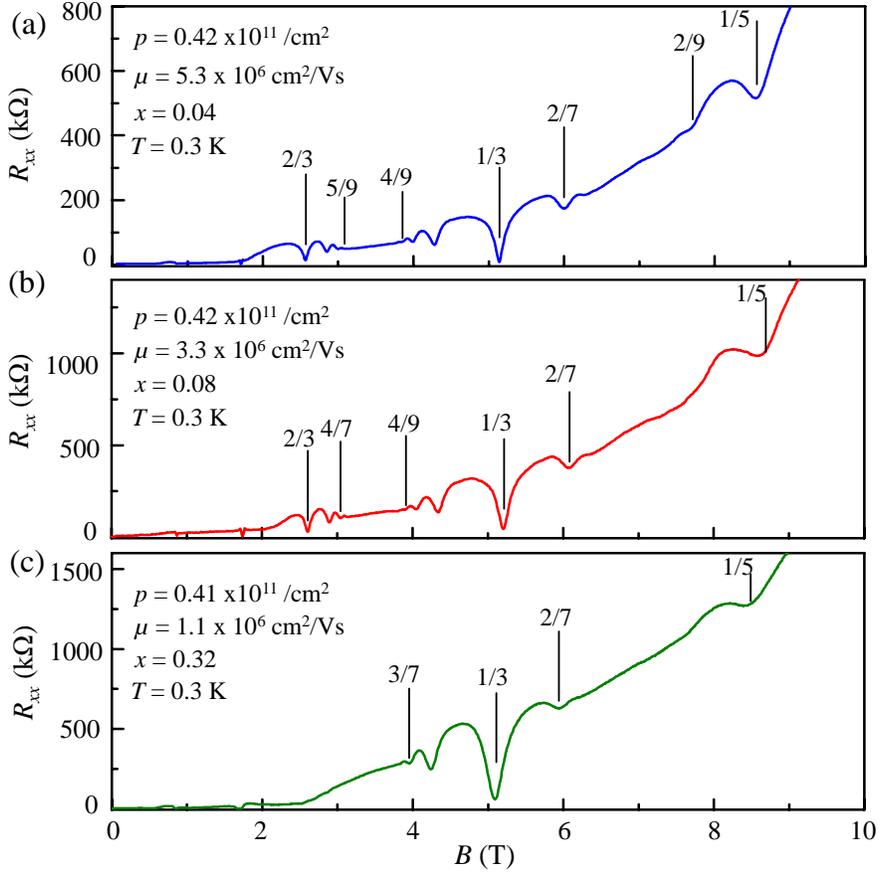

Fig. S3. Comparison of magnetoresistance data at a low density. Representative $R_{xx}$ vs $B$ traces for 2DHSs for: (a) $x = 0.04$, (b) $x = 0.08$, and (c) $x = 0.32$ at a 2DHS density $p \simeq 0.4 \times 10^{11}$ /cm$^2$. The three samples have $w = 20$ nm, and all the traces are taken at $T = 0.3$ K. The magnetic field positions of some fractional quantum Hall states are marked.

In this section we provide additional evidence for the enhancement in quality of our samples by comparing the magnetoresistance data at $T = 0.3$ K for samples with different barrier alloy fractions near the GaAs QW. Figure S3 compares the $R_{xx}$ vs $B$ traces at $p \simeq 0.4 \times 10^{11}$ /cm$^2$ for $w = 20$ nm samples with different $x$. Remarkably, even at 0.3 K, all samples exhibit clear signs of developing four-flux fractional quantum Hall states (FQHSs) at $\nu = 1/5$ and 2/7, with the strength of these states weakening with increasing $x$. The $x = 0.04$ sample additionally shows an inflection point at $\nu = 2/9$. Clearly, the $x = 0.32$ sample has inferior quality than both $x = 0.08$ and $x = 0.04$ samples as evinced by the weaker and smaller number of FQHS minima near $\nu = 1/2$. The other two samples both show excellent quality, with slightly stronger FQHSs in the $x = 0.04$

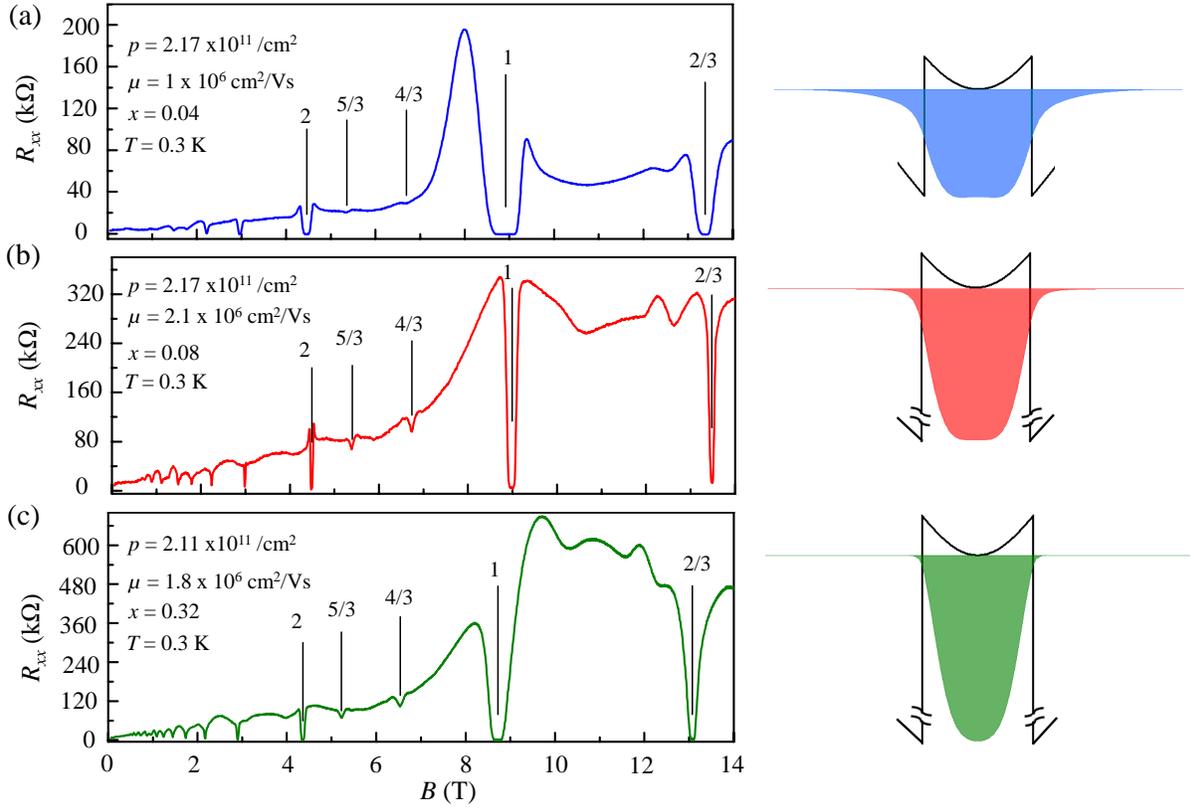

Fig. S4. Comparison of magnetoresistance data at a high density. Representative $R_{xx}$ vs $B$ traces at 0.3 K for 2DHSs for: (a) $x = 0.04$, (b) $x = 0.08$, and (c) $x = 0.32$ at a 2DHS density $p \simeq 2.2 \times 10^{11}$ /cm$^2$. For all three samples, $w = 20$nm. The magnetic field positions of some integer and fractional quantum Hall states are marked. The self-consistently calculated hole charge distribution and potential functions are shown for each case.

sample.

In Fig. S4, we compare traces at a higher density $p \simeq 2.2 \times 10^{11}$ /cm$^2$; all samples have $w = 20$ nm. At this density, alloy disorder scattering limits the mobility as discussed in the last section. The self-consistent charge distributions clearly depict the tails extending into the barrier regions with maximum penetration for $x = 0.04$. Consequently, we observe the lowest mobility for $x = 0.04$ in contrast to low densities (Fig. S3). The same is reflected in the magnetoresistance trace with broad integer and fractional QHSs. At this density, $x = 0.08$ shows the best mobility and quality with sharp minima at both integer and fractional states.

Figures S3 and S4 demonstrate that mobility is indeed strongly correlated with quality. The comparison between the $x = 0.08$ and $x = 0.32$ data in Fig. S4 also indicates that, even though the penetration of the wavefunction into the barriers is higher for $x = 0.08$ than $x = 0.32$, the alloy



\begin{thebibliography}{70}
\bibitem{footnote1}The calculated $m^*$ is approximate because it assumes same mass for both spin-split subbands. Limited measurements of $m^*$ using low-field Shubnikov de Haas oscillations suggest that calculations overestimate m* by $\sim$30\%. However, this does not affect our mobility analysis (Fig.~\ref{fig:fig2}) significantly because $m^*$ also enters in Eq.~\ref{eq:tau} through dielectric screening (Eq.~\ref{eq:epsilon}), and therefore reduces the dependence of mobility on $m^*$.
\bibitem{footnote3} It is also possible that screening affects individual scattering mechanisms differently, which may further affect the temperature dependence of each scattering mechanism and the resultant temperature dependence of mobility. More theoretical work on this front could be useful for discerning individual contributions of scattering mechanisms to the temperature dependence of mobility.
\end{thebibliography}
\end{document}